# Did Einstein "Nostrify" Hilbert's Final Form of the Field Equations for General Relativity?

Galina Weinstein

December 5, 2014

Tel-Aviv University

Einstein's biographer Albrecht Fölsing explained: Einstein presented his field equations on November 25, 1915, but six days earlier, on November 20, Hilbert had derived the identical field equations for which Einstein had been searching such a long time. On November 18 Hilbert had sent Einstein a letter with a certain draft, and Fölsing asked about this possible draft: "Could Einstein, casting his eye over this paper, have discovered the term which was still lacking in his own equations, and thus 'nostrified' Hilbert?" Historical evidence support a scenario according to which Einstein discovered his final field equations by "casting his eye over" his own previous works. In November 4, 1915 Einstein wrote the components of the gravitational field and showed that a material point in a gravitational field moves on a geodesic line in space-time, the equation of which is written in terms of the Christoffel symbols. Einstein found it advantageous to use for the components of the gravitational field the Christoffel symbols. Einstein had already basically possessed the field equations in 1912, but had not recognized the formal importance of the Christoffel symbols as the components of the gravitational field. Einstein probably found the final form of the generally covariant field equations by manipulating his own (November 4, 1915) equations. Other historians' findings seem to support the scenario according to which Einstein did not "nostrify" Hilbert.

David Hilbert invited Einstein to come to Göttingen to talk about his new 1914 gravitation theory that is presented in the paper, "The Formal Foundation of the General Theory of Relativity".[1] Einstein held the lectures under the auspices of the Wolfskehl Foundation in Göttingen the week beginning Monday, **28 June** and he returned to Berlin on **July 5, 1915**.[2] All that is left from these lectures is 11 pages of notes by an unknown auditor of part of these lectures. In this *Nachschrift* the unknown auditor reported about Einstein's talk on "The gravitational Field and the Equation of Motion of a material Point in this Field".[3]

On **July 7, 1915** Einstein wrote Heinrich Zangger about his impression of Hilbert: "I got to know and love him. I held six two-hour lectures there on the gravitation theory, which is now clarified very much, and had the experience of convincing the mathematician friends there completely".[4] A couple of weeks later, Einstein wrote him again: "I have probably written you that I held 6 lectures at Göttingen, where I was able to convince Hilbert of the general theory of relativity. I am enchanted with the latter, a man of admirable energy and independent personality in all things. Sommerfeld is also beginning to agree; Planck and Laue stay aloof".[5] Einstein was mostly satisfied: "at the end of June-beginning of July I held six detailed lectures at Göttingen on general relativity theory. To my great satisfaction I succeeded in convincing Hilbert and [Felix] Klein completely".[6] At the same time Hilbert apparently was also enchanted with Einstein and his new 1914 general relativity theory.

In his first November 1915 paper on General Relativity, the November 4 1915 paper, Einstein gradually expanded the range of the covariance of his gravitation field equations. Every week he expanded the covariance a little further until on November 25 he arrived to fully generally covariant field equations.

In the introduction to the **November 4, 1915** paper Einstein said he believed he had found the only law of gravitation that complies with a reasonably formulated postulate of general relativity. He had already tried to demonstrate the truth of this solution, in the 1914 paper.[7] But he "completely lost trust in my established field equations, and looked for a way to limit the possibilities in a natural manner. Thus I arrived back at the demand of a broader general covariance for the field equations, from which I parted, though with a heavy heart, three years ago when I worked together with my friend [Marcel] Grossmann. As a matter of fact, we then have already come quite close to the solution of the problem given in the following". Just as the special theory of relativity is based upon the postulate that all equations have to be covariant relative to linear orthogonal transformations, so the theory developed by Einstein in the November 4 paper rests upon the postulate of the covariance of all systems of equations relative to transformations with the substitution determinant 1.[8]

On **November 7, 1915** Einstein wrote David Hilbert and sent him the proofs of the November 4 paper: "I am sending you the correction to a paper in which I changed the gravitation equations". Einstein told Hilbert that he had corrected his 1914 review paper into the November 4, 1915 paper, "after I myself realized about 4 weeks ago that my previous method of proof was fallacious". Since Hilbert had "found a hair in my soup, which spoiled it entirely for you" (a mistake in Einstein's 1914 paper), Einstein wished him to look at his new work.[9] Later on March 30, 1916, Einstein sent a letter to Hilbert in which he explained the mistake:[10] "The error you found in my paper of 1914 has now become completely clear to me".[11]

By **November 10, 1915** Hilbert probably answered Einstein's letter, telling him about his system of electromagnetic theory of matter, the unified theory of gravitation and electromagnetism, in which the source of the gravitational field is the electromagnetic field. Hilbert's goal was to develop an electromagnetic theory of matter, which would explain the stability of the electron.[12]

A week later, the mathematical-physical class of the Prussian academy gathered again to hear a correction to Einstein's November 4 paper. Einstein presented this correction in his Addendum, "Zur allgemeinen Relativitätstheorie (Nachtrag)".[13] He dropped his November 4 postulate and adopted it as a coordinate condition that allowed him to take the last step in order to write the field equations of gravitation in a general covariant form.[14]

Einstein had only waited one day and on **November 12, 1915** he wrote his new friend Hilbert from Göttingen. He first thanked him for his kind letter a reply, sent sometime **between November 8 and 10** to Einstein's letter from November 7.[15] Perhaps Hilbert told Einstein: your equations from the November 4, 1915 paper needed reconsideration. Hilbert's letter was not survived and we could only speculate about the contents of the letter. Einstein wrote to Hilbert on November 12: "the problem has meanwhile made new progress. Namely, it is possible to exact general covariance from the postulate $\sqrt{-g} = 1$; Riemann's tensor then delivers the gravitation equations directly".[16] Hence, Einstein reported about a progress and a main finding regarding $\sqrt{-g} = 1$, to which he very likely arrived as a result of reconsideration of his equations from the November 4, 1915 paper.

Nonetheless it is certainly reasonable that when Einstein arrived at the above solution he was influenced by Hilbert's possible letter that had been sent **before November 10**. In the addendum, the November 11 paper, Einstein added the following coordinate condition: "*we assume in the following that the condition* $\sum T_\mu^\mu = 0$ *is in general actually fulfilled* ".[17] Einstein had noticed that this condition, which follows from setting $\sqrt{-g} = 1$, can be related to an electromagnetic theory of matter.[18] Einstein arrived at the above hypothesis by making new assumptions on electromagnetic and gravitational "matter".

What impelled Einstein to change his perspective in the November 11 paper? Jürgen Renn and John Stachel say that it would have been quite uncharacteristic of Einstein to adopt the new approach so readily had it not been for current discussions of the electrodynamic worldview and his feeling that he was now in competition with Hilbert. When one examines Einstein's previous writings on gravitation, one finds no trace of an attempt to unify gravitation and electromagnetism. He had never advocated the electromagnetic worldview. On the contrary, he was apparently disinterested in attempts at a unification of gravitation and electrodynamics. Moreover, soon after completion of the final version of general relativity (circa November 25), Einstein reverted to his earlier view that general relativity could make no assertions about the structure of matter. Einstein's mid-November 1915 pursuit of a relation between gravitation and electromagnetism was, then, merely a short-lived episode in his search for a relativistic theory of gravitation.[19]

Einstein received a very prompt response from Hilbert. On **November 13, 1915** Hilbert replied to Einstein's letter. He had already been fully immersed in Einstein's problem: "Actually, I first wanted to think of a very palpable application for physicists, namely reliable relations between the physical constants, before obliging with my axiomatic solution to your great problem. But since you are so interested, then this coming Tuesday I will develop, approximately after tomorrow (d. 16 of M.), my whole theory in detail". In the letter he noted the main points of his theory, a unification of gravitation and electromagnetism and told Einstein that he had already discussed his discovery with Arnold Sommerfeld. He wanted to explain it also to Einstein on Tuesday and invited him to come at three or half past five. "The Math. Soc. shall meet at 6 o'clock in the auditorium-house. My wife and I would be very pleased if you stayed with us". Hilbert, however, appended a note to his letter: "As far as I understand your new pap. [paper of November 4], the solution giv. [given] by you is entirely different from mine […]", and he sent Einstein an extra page: "Continuation on page I with the invitation to come for Tuesday, 6 o'clock, many greetings H".[20] Indeed on Tuesday, **November 16**, Hilbert presented his talk "Grundgleichungen der Physik" ("The Fundamental Equations of Physics") to the Mathematical Society of Göttingen.[21]

On **November 15, 1915** Einstein replied to Hilbert. He showed little sign of being willing to come at the moment to Göttingen. At any rate, Einstein was enormously curious about Hilbert's findings but things have come downhill emotionally: "should I wait until I can study your system from the printed paper, for I am very tired and in addition plagued with stomach pains. Please send me if possible, a proof copy of your investigation in order to satisfy my impatience".[22]

A reliable quick replier to letters Hilbert was! On **November 16**, Hilbert perhaps sent a copy of the lecture he had given on the subject or else a copy of a manuscript of the paper he would present five days later on **November 20** to the Royal Society in Göttingen.[23]

It is evident that until **November 17** Einstein had been still in competition with Hilbert and was influenced by his electromagnetic theory of matter. Indeed Einstein told his close friend Michele Besso, "In these last months, I had great success in my work. *Generally covariant* gravitation equations. *Perihelion motions explained quantitatively*. The role of gravitation in the structure of matter".[24]

Renn and Stachel examined Einstein's correspondence with Hilbert and found it is quite clear that this perspective of a theory of matter was shaped by Einstein's rivalry with Hilbert. It thus seems evident that Einstein's temporary adherence to an electromagnetic theory of matter was triggered by Hilbert's work, which Einstein attempted to use in order to solve a problem that had arisen in his own theory, and that he dropped it when he solved this problem in a different way.[25]

At this stage of the game, Einstein was already less patient after he had received Hilbert's "System". On **November 18, 1915** he replied to Hilbert telling him that "Your given system agrees – as far as I can see – exactly with what I found in the last few weeks and have presented to the Academy".[26] Hilbert probably had put enormous effort of work and thought had gone into his system, but according to Einstein he did not reinvent the wheel. Hence, Hilbert's system of equations (mentioned in Hilbert's November 13 letter) agreed with Einstein's November 4 equations and in the November 7 letter Einstein had perhaps sent him these equations. But Hilbert did not think so; on **November 13** he wrote to Einstein: "As far as I understand your new pap. [paper of November 4], the solution giv. [given] by you is entirely different from mine […]". Leo Corry, Renn, and Stachel say that "it is clear from this exchange that, at this point, at least one of the two – and probably each – is misunderstanding the other's work".[27]

In the **November 18** letter to Hilbert Einstein explained:

"The difficulty was not in finding generally covariant equations for the $g_{\mu\nu}$'s; for this is easily achieved with the aid of Riemann's tensor. Rather, it was hard to recognize that these equations are a generalization, that is, a simple and natural generalization of Newton's law. I did it only in the last few weeks (my first communication that I sent you [November, 4]), whereas I considered the only possible generally covariant equations, which have now been proven to be correct, 3 years ago with my friend Grossmann [Zurich Notebook]. With heavy heart we separated from them, because it seemed to me that the physical discussion revealed an incompatibility with Newton's law. – The important thing is that the difficulties have now been overcome. I am presenting today to the Academy a paper in which I derive quantitatively out of general relativity, without any guiding hypothesis, the perihelion motion of Mercury discovered by Le Verrier. This was not achieved until now by any gravitational theory".[28]

The day afterwards, on **November 19, 1915** Hilbert sent a polite letter in which he congratulated Einstein "on overcoming the perihelion motion. If I could calculate as rapidly as you […]".[29] In fact Einstein did not calculate the result that rapidly. A week after November 11, on the coming Thursday, he presented his work on the Perihelion Motion of Mercury; but the basic calculation was done two years earlier with Michele Besso in the Einstein-Besso manuscript. Einstein transferred the basic framework of the calculation from the Einstein-Besso manuscript and corrected it according to his November 11 field equations.[30]

Although Einstein rapidly calculated the perihelion shift of Mercury on the basis of his previous work with Besso, the success of his calculation was also based on his November 11 theory. The condition $\sqrt{-g} = 1$, implied by the assumption of an electromagnetic origin of matter, was essential for this calculation, which Einstein considered a striking confirmation of his audacious hypothesis on the constitution of matter, definitely favoring this theory over that of November 4. Thus when writing the Perihelion paper Einstein was still influenced by Hilbert's electromagnetic theory of matter.[31] On the other hand, Einstein was already departing from Hilbert's theory of matter in his November 18 work, because he wrote, "through which time and space are deprived of the last trace of objective reality".[32] Hence, Einstein began to realize that space and time coordinates have no meaning in general relativity.

Einstein had very busy weeks when working on the problem of the Perihelion of Mercury. Sometime between November 11 and November 25 he could resolve the final difficulties in his theory. It probably took him just a couple of days to arrive at the November 25 field equations. He expanded his November 11 field equations and presented the final version of the field equations. We can ask questions of priority nature: When did Einstein get the idea to expand his November 11 equations, before he had received Hilbert's "System" or afterwards? Did Hilbert's "System" induce Einstein to arrive at the final solution of November 25?

Hilbert ended his **November 19, 1915** letter by asking Einstein to continue and keep him up to date on his latest advances but, he did not tell Einstein about a particularly important talk he planned to give the day afterwards. Hilbert presented on **November 20** a paper to the Göttingen Academy of Sciences, "The Foundations of Physics", including his version to the gravitational field equations of general relativity. Five days later **on November 25**, Einstein presented to the Prussian Academy his version to the gravitational field equations.

On **November 26, 1915** a day after Einstein presented the final version of the field equations he wrote his close friend Zangger:[33]

"The general relativity problem is now finally dealt with. The perihelion motion of Mercury is explained wonderfully by the theory. […] The theory is beautiful beyond comparison. However, only *one* colleague has really understood it, and he is seeking to clearly "nostrify" it (Abraham's expression). In my personal experience I have hardly come to know the wretchedness of mankind sometimes better than this theory and everything connected to it. But it does not bother me".[34]

In 1912, Max Abraham had blamed Einstein's theory of relativity and Einstein as well. Abraham thought that Einstein borrowed expressions from his new gravitation theory. On March 26, 1912, Einstein sent his best friend Michele Besso a description of his most exciting development regarding the static gravitational field and did not forget Abraham: "Abraham's theory was created of the top of his head, i.e., from mere mathematical beauty considerations, torn off and completely untenable".[35] That was almost Abraham's opinion of Einstein's theory, except for the mathematical beauty.

Abraham formulated his theory in terms of Minkowski's four-dimensional space-time formalism and Einstein's 1911 relation between the variable velocity of light and the gravitational potential.[36] Einstein attacked Abraham's theory because of what he considered the incompatibility between Abraham's simultaneous implementation of both a variable speed of light and Minkowski's space-time formalism. He also thought that Abraham's theory could not be reconciled with a theory that was based on the equivalence principle. Abraham gradually converted to elements of Einstein's theory.

In 1912 Abraham criticized Einstein for not yet adopting Minkowski's space-time formalism and he did not like Einstein's use of the equivalence principle.

Abraham, however, needed Einstein's result of the mass of energy principle for his theory of gravitation. It was Abraham who corrected his theory according to Einstein's ideas, and not the other way round. After blaming Einstein, however, and because of the resentment he had toward the theory of relativity, Abraham could not accept this state of affairs. He thus found an original solution to the particularly important question: who actually arrived at the idea of the mass of energy? On August 16, 1912, Einstein explained this unpleasant experience to Ludwig Hopf: "Recently, Abraham – as you may have seen – slaughtered me along with the theory of relativity in two massive attacks, and wrote down (phys. Zeitschr.) the only correct theory of gravitation (under the 'nostrification' of my results) – a stately steed, that lacks three legs! He noted that the knowledge of the mass of energy comes from – Robert Mayer".[37]

Hence, "nostrification" was Einstein's expression and not Abraham's but, the mass of energy indeed came from Einstein. We can, however, readily admit that history repeats itself. After the polemic with Abraham, Einstein encountered Gunnar Nördstrom: competition, and so forth. Nördstrom similarly to Abraham also gradually converted to Einstein's theory of gravitation. Nördstrom needed Einstein's corrections to his theory, his equivalence principle and also Einstein's help with conservation of energy. Competition, however, is not "root of all evil", because Einstein also needed the interaction with his competitors: Nördstrom and Abraham at some stage had inspired Einstein.

In 1912, in Zurich, Einstein understood that Minkowski's formalism was crucial for the further development of his theory of gravitation and he applied it once he had recognized that the gravitational field should be described by the metric tensor field. In Zurich Einstein tried to solve the field equations of Nordström's 1913 theory. One page in the Einstein-Besso manuscript of 1913, page 53, contains calculations of the Perihelion motion of an orbit in the weak field of a static sun on the basis of Nordström's theory of gravitation. This page discusses how the gravitational field of the sun is found in the Nordström theory, and how the Perihelion motion of an orbit in this field is found.[38] Finally, in September 1913, Einstein examined a competing theory to his own gravitation theory that was based on Nordström's theory.[39]

In 1915 Hilbert worked on the same problem as Einstein and needed Einstein's November 4 and 1914 papers for his theory of gravitation. It appears that Einstein's papers were crucial for the development of his theory but Hilbert also inspired Einstein in his November 11 and 18 works. According to Einstein, however, Hilbert was "seeking to clearly 'nostrify'" his theory. Did Hilbert's "System" induce Einstein to arrive at the final solution of November 25?

On Thursday November 25 Einstein presented to the mathematical physical class of the Prussian Academy of sciences his short paper "Die Feldgleichungen der Gravitation" (The Field Equations of Gravitation"). What happened between November 18 and 25 that brought Einstein back to the Prussian Academy with the final version of his field equations? Einstein wrote, "I now quite recently found that one can get along without this hypothesis about the energy tensor of matter, merely by inserting it into the field equations in a slightly different way. The field equations for vacuum, onto which I based the explanation of the Mercury perihelion, remain unaffected by this modification". He briefly summarized his previous two papers of November 4 and 11 and historically explained the way his field equations evolved. Finally, he wrote the new field equations of November 25. [40]

Einstein's biographer Albrecht Fölsing wrote: "Der 'nostrifizierende' Kollege war ausgerechnet David Hilbert, von dem er im Sommer noch 'ganz begeistert' war aber noch mehr dürfte ihn geärgert haben, daß Hilbert sogar schon einige Tage früher als er selbst die richtigen Feldgleichungen veröffentlicht hatte". [41]

Fölsing explained that the "nostrified" colleague ['nostrifizierende' Kollege] was Hilbert and he was excited in the summer, a few days before Einstein had published the November 25 field equations. Einstein presented his field equations on November 25, 1915, but six days earlier, on November 20, Hilbert had derived the identical field equations for which Einstein had been searching such a long time. Fölsing asked about a possible draft that Hilbert sent Einstein before November 18: "Hat Einstein beim Uberfliegen dieser Arbeit denjenigen Term entdeckt, der in seinen Gleichungen noch fehlte, und dadurch etwa Hilbert 'nostrifiziert'?" Could Einstein, casting his eye over this paper, have discovered the term which was still lacking in his own equations, and thus 'nostrified' Hilbert?" [42]

Fölsing took Einstein's phraseology "nostrifiziert" and turned it against him. By apparent contrast, historical evidence support a scenario according to which Einstein discovered the second term in his November 25 field equations by "casting his eye over" his own works of November 4, 1915 and 1914.

On November 28, 1915, Einstein explained to Arnold Sommerfeld: [43]

"The key to this solution was my realization that […] the related Christoffel symbols […] are to be regarded as the natural expression of the gravitational field 'components'. Once one sees this, then the above equation is very simple, there is no need to transform it for the purpose of a general interpretation by computing the symbols".

On January 1, 1916, he wrote to Hendrik Antoon Lorentz: [44]

"I had already basically possessed the current equations 3 years ago together with Grossman, who had brought my attention to the Riemann tensor. But because I had not recognized the formal importance of the { } terms [Christoffel symbols], I could not obtain a clear overview and prove the conservation laws […]. Then I fell into the jungle!"

Hence, in **October-November 1914** Einstein had written the components of the gravitational field and had shown that a material point in a gravitational field moves on a geodesic line in space-time, the equation of which is written in terms of the Christoffel symbols. By **November 4, 1915**, Einstein found it advantageous to use for the components of the gravitational field the Christoffel symbols. Einstein had already basically possessed the field equations **in 1912** together with Grossman, but because he had not recognized the formal importance of the Christoffel symbols as the components of the gravitational field, he had fallen "into the jungle". Finally, Einstein probably found the final form of the generally covariant field equations of **November 25, 1915** by manipulating his own (November 4, 1915) equations.

In his November 4, 1915 paper Einstein wrote an equation of conservation of energy-momentum: [45]

$$(1) \sum_\nu \frac{\partial T_\sigma^\nu}{\partial x_\nu} = \frac{1}{2} \sum_{\mu\tau\nu} g^{\tau\mu} \frac{\partial g_{\mu\nu}}{\partial x_\sigma} T_\tau^\nu,$$

($g_{\mu\nu}$ is the metric tensor and $T_\sigma^\nu$ denotes the "energy tensor") He noted: [46]

"This equation of conservation led me in the past to view the quantities

$$\frac{1}{2} \sum_\mu g^{\tau\mu} \frac{\partial g_{\mu\nu}}{\partial x_\sigma}$$

as the natural expression of the components of the gravitational field, even though in view of the formulas of the absolute differential calculus, it is better to introduce the Christoffel symbols

$$\begin{Bmatrix} \nu\sigma \\ \tau \end{Bmatrix}$$

instead of these quantities".

In his November 25, 1915 paper Einstein wrote the field equations in the following form: [47]

$$(2) \; R_{im} = -\kappa \left( T_{im} - \frac{1}{2} g_{im} T \right).$$

$$(3) \sum_{\rho\sigma} g^{\rho\sigma} T_{\rho\sigma} = \sum_\sigma T_\sigma^\sigma = T$$

T is the "scalar" of the momentum-energy tensor (for a frame of reference $\sqrt{-g}$ = **1**).

Perhaps the "equation of conservation" (1) led Einstein to view the quantities $\frac{1}{2}\sum_{\mu\tau\nu} g^{\tau\mu} \frac{\partial g_{\mu\nu}}{\partial x_\sigma} T^\nu_\tau$ as a natural expression of the final form of the field equations. This term on the right-hand side of equation (1) seemingly could, however, be a heuristic guide for Einstein for the second term on the right-hand side (1/2 $g_{im}$T) of equation (2). In contrast to Hilbert, Einstein searched for a gravitational field equation that would satisfy some heuristic requirements.

Einstein was guided by several heuristic principles: the principle of relativity, the equivalence principle, the correspondence principle, and the principles of conservation of energy and momentum. The energy-momentum conservation principle played a crucial role in Einstein's General Theory of Relativity and he required that the gravitational field equation should be compatible with the generalized requirement of energy and momentum conservation. [48]

Findings of other historians regarding Hilbert's November 20 paper further support my arguments and findings described above according to which Einstein very likely did not take anything pertaining to the November 25 field equations from Hilbert's paper, or the summary sent to him.

Indeed, the situation was quite the reverse: the question is whether Hilbert might have taken something from Einstein's November 25 paper, because it was quite possible for him to do so.[49] Hilbert submitted his paper to the Göttingen Academy of Sciences on **November 20, 1915**. Einstein's paper in which he gave the final form of his generally covariant field equations was submitted to the Prussian Academy of Sciences on **November 25, 1915**. Hilbert very likely sent to Einstein (on November 16?) **before November 18** a summary of his November 20 work.[50]

The **November 20, 1915** proofs bear a printer's date stamp, "**6 December 1915**". The paper was not yet published.[51] According to Fölsing, the November 20, 1915 proofs of Hilbert paper are equivalent to Hilbert's printed paper and thus contain an equivalent version of Einstein's November 25 field equations. The November 20, 1915 paper, however, did not contain a generally covariant theory. And thus represent Hilbert's states of work submitted on November 20. Hilbert started correcting his proofs only on December 6, 1915.

Einstein's **November 25, 1915** paper was published on **December 2, 1915**. Hilbert's paper was published only on **March 31, 1916**, and he had plenty of time to correct his November 20, 1915 paper according to Einstein's published work of **December 2, 1915**. Hilbert indeed rewrote his November 20 paper sometime between December 1915 and March 1916.[52]

There are, however, differences between the November 20 paper and the printed version from March 1916. In the November 20 proofs Hilbert based his assertion on a slightly more sophisticated version of Einstein's Hole Argument against general covariance (after Einstein had silently dropped it), which he would eventually drop later when he would publish his paper in March 1916.[53] In addition, in Hilbert's proofs of November 20 the gravitational field equations do not appear explicitly. In the published version of March 1916 – after Einstein had published the final form of his field equations – the expression equivalent in form to Einstein's November 25 field equations is written down explicitly. Corry, Renn and Stachel indeed claim that knowledge of Einstein's result may have been crucial to Hilbert's introduction of the second term in his equation, which was equivalent in form to these equations of Einstein. And thus Einstein's papers helped Hilbert in putting his November 20 paper in a malleable form containing generally covariant field equations. Finally, Hilbert later supplemented his reference to the gravitational potentials $g_{\mu\nu}$ in handwriting with the phrase "first introduced by Einstein".[54]

On **December 20, 1915** Einstein wrote Hilbert,[55]

"There has been certain resentment between us, the cause of which I do not want to analyze. I have fought against the associated feeling of bitterness with complete success. I think of you again with unmixed kindness, and I ask you to try to do the same with me. It is objectively a shame when two real guys that have emerged from this shabby world do not give each other a little pleasure".

In the March 1916 printed version of his November 20 paper, Hilbert added a reference to Einstein's November 25 paper and wrote: "the differential equations of gravitation that result are, as it seems to me, in agreement with the magnificent theory of general relativity established by Einstein in his last papers".[56]

On **May, 27, 1916** Hilbert invited Einstein to visit Göttingen again and stay with him; but in spite of several invitations over the next few years, Einstein never came, but they continued to correspond over issues connected with Hilbert's paper.

---

[1] Einstein, Albert (1914). "Die formale Grundlage der allgemeinen Relativitätstheorie." *Königlich Preußische Akademie der Wissenschaften* (Berlin). *Sitzungsberichte*, 1030-1085.

[2] *The Collected Papers of Albert Einstein. Vol. 8: The Berlin Years: Correspondence, 1914–1918* (*CPAE* 8), Schulmann, Robert, Kox, A.J., Janssen, Michel, Illy, Jószef (eds.), Princeton: Princeton University Press, 2002, note 5, 146.

[3] *The Collected Papers of Albert Einstein. Vol. 6: The Berlin Years:Writings, 1914–1917* (*CPAE* 6), Klein, Martin J., Kox, A.J., and Schulmann, Robert (eds.), Princeton:

Princeton University Press, 1996, Appendix B, "*Nachschrift* of Einstein's Wolfskehl lectures, Summer 1915", 589.

[4] Einstein to Zangger, July 7, 1915, *CPAE* 8, Doc. 94.

[5] Einstein to Zangger, Between July 24 and August 7, 1915, *CPAE* 8, Doc. 101.

[6] Einstien to Wander and Geertruida de Hass, August 16, 1915, *CPAE* 8, Doc. 110.

[7] Einstein 1914, sections §12-§15, 1066-1077.

[8] Einstein, Albert (1915a). "Zur allgemeinen Relativitätstheorie." *Königlich Preußische*, *Akademie der Wissenschaften* (Berlin). *Sitzungsberichte*, 778-786; 778.

[9] Einstein to Hilbert, November 7, 1915, *CPAE* 8, Doc. 136.

[10] *CPAE* 8, note 4, 192.

[11] Einstein to Hilbert, March 30 1916, *CPAE* 8, Doc. 207.

[12] Stachel, John (2005). "Einstein and Hilbert." Invited lecture in March 21, 2005 Session: Einstein and Friends, American Physical Society, Los Angeles; and response to Questions from FAZ on Hilbert and Einstein".

[13] Einstein, Albert (1915b). "Zur allgemeinen Relativitätstheorie. (Nachtrag)." *Königlich Preußische Akademie der Wissenschaften* (Berlin). *Sitzungsberichte*, 799-801.

[14] Einstein, 1915b, 800.

[15] Einstein to Hilbert, November 12, 1915, *CPAE* 8, Doc. 139.

[16] Einstein to Hilbert, November 12, 1915, *CPAE* 8, Doc. 139.

[17] Einstein, 1915b, 799-800.

[18] Renn, Jürgen and Stachel, John (2007). "Hilbert's Foundation of Physics: From a Theory of Everything to a Constituent of General Relativity." In Renn, Jürgen, Norton, John, Janssen, Michel and Stachel John, ed. (2007). *The Genesis of General Relativity*. 4 Vols., New York, Berlin: Springer, 857-974; 904.

[19] Renn and Stachel 2007, 905-907.

[20] Hilbert to Einstein, November 13, 1915, *CPAE* 8, Doc. 140.

[21] *CPAE* 8, note 2, 196.

[22] Einstein to Hilbert, November 15, 1915, *CPAE* 8, Doc. 144.

[23] *CPAE* 8, note 1, 202.

[42] Fölsing, 1993, 421; Fölsing, Albrecht (1997). *Albert Einstein, A Biography*. Translation by Ewald Osers, New York: Penguin books, 375-376.

[43] Einstein to Sommerfeld, November 28, 1915, *CPAE* 8, Doc. 153.

[44] Einstein to Lorentz, January, 1 1916, *CPAE* 8, Doc. 177.

[45] Einstein 1915a, 782, 784. For a discussion and an explanation see my paper: Weinstein, Galina, "Einstein and the conservation of energy-momentum in general relativity", ArXiv: 1310.2890v [physics.hist-ph], 11 Oct, 2013.

Einstein, 1915a, pp. 782-783. [46]

[47] Einstein 1915d, 845.

[48] Renn, Jürgen and Sauer, Tilman (2007). "Pathways out of Classical Physics. Einstein's Double Strategy in his Search for the Gravitational Field Equation." In Janssen, Norton, Renn, Sauer, Stachel (2007), 113-312; 123-125, 147.

[49] Stachel, John (1999). "New Light on the Einstein-Hilbert Priority Question." *Journal of Astrophysics* 20, 90-91. Reprinted in Stachel (2002). *Einstein from 'B' to 'Z'*, Washington D.C.: Birkhauser, 353-364; 357.

[50] Stachel 1999, 358.

[51] Corry, Leo, Renn, Jürgen and John Stachel (1997). "Belated Decision in the Hilbert-Einstein Priority Dispute." In Stachel 2002, 339-346; 340.

[52] Stachel 1999, 358.

[53] Stachel 1999, 359.

[54] Corry, Renn and Stachel 1997, 343, 359.

[55] Einstein to Hilbert, December 20, 1915, *CPAE* 8, Doc. 167.

[56] Corry, Renn and Stachel 1997, 344.